# DESIGNING AND IMPLEMENTING THE LOGICAL SECURITY FRAMEWORK FOR E-COMMERCE BASED ON SERVICE ORIENTED ARCHITECTURE

Ashish Kr. Luhach[1], Dr. Sanjay K. Dwivedi[2], Dr. C. K. Jha[3]

[1]Dronacharya College of Engineering, Gurgaon, Hr, India
[2]BBA University, Lucknow, U.P., India
[3]Banasthali University, Jaipur, Rajasthan, India.

*ABSTRACT*

*Rapid evolution of information technology has contributed to the evolution of more sophisticated E-commerce system with the better transaction time and protection. The currently used E-commerce models lack in quality properties such as logical security because of their poor designing and to face the highly equipped and trained intruders. This editorial proposed a security framework for small and medium sized E-commerce, based on service oriented architecture and gives an analysis of the eminent security attacks which can be averted. The proposed security framework will be implemented and validated on an open source E-commerce, and the results achieved so far are also presented.*

*KEYWORDS*

*Web services, E-commerce, Simple Object Access Protocol*

## 1. INTRODUCTION

With the evolution in information engineering, internet users increased exponentially and as a consequence of which websites are utilized as the new and integrated marketing. These websites eliminated the physical part of markets. Websites provide a virtual place for the user to perform online transactions, known as E-markets. These E-markets are becoming the core of attraction for researchers and initiatives. E-commerce systems are mainly based on networks and computing power of the users. At present, the problem exists in E-Commerce's are due to frequent update of business process because of changing customer demands and platform integration because of the heterogeneous platform used by different enterprises. To subdue the alleged problems in business operation and integration, Service Oriented Architecture (SOA) can be applied.  SOA is an information technology approach in which the existing applications in an enterprise can use the various services available in a network i.e. World Wide Web.

With the help of SOA, such applications can be evolved, as well. SOA also has the feature of loose coupling, which resolves the problem of transmission and integration of information. SOA has an open standard protocols and excellent encapsulation, which makes SOA the suitable choice to implement E-commerce.

## 2. E-COMMERCE BASED ON SOA

In general, the most important aspect in business from the customer's point of view is cost. E-commerce systems need techniques or technical solutions, which reduces the processing or transaction cost as the number of users increased exponentially in recent years for performing online transactions [2]. Another important problem to address while discussing E-commerce's is Interoperability because E-commerce's may vary from each other in terms of business flow and data description. This variation depends upon communication protocols, and object models used.

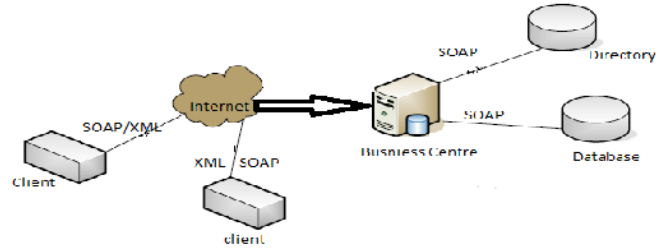

Figure1: E-commerce based on SOA

As a consequence of which, if any one of the partner involved in the E - commerce scenario made any alterations in their realization mechanism, then to adopt these changes other partners also have to make modifications in their operation. Debating the future of E-commerce development, Interoperability is still an unresolved inquiry. At present, a great deal of research and conferences are conducted on E-commerce just to furnish backup to the enterprises for their migration towards E-commerce. However, Service Oriented Architecture (SOA) proves to be one of the best technological results to implement E-commerce. The general idea behind developing SOA is to integrate software into services. SOA is a distributed computing technology, which provides services through XML protocols. SOA applications embedded into network services using XML. Web Services attracts enterprises and business to get the benefits of connectivity through the networks [2]. SOA can help organizations to recycle their existing assets such as programming code with significant changes and produce a solution for application consolidation, which saves cost and execution time.SOA provides a reliable platform for communication and integration between E-commerce system and their associated partners; this will explore new dimensions of the enterprise. The principal characteristics of SOA are Interoperability and Reusability. However to solve the above discussed problems in E-commerce, SOA proves to be the best solution for implementing E-commerce. Figure 1 show an E-commerce based on SOA.

## 3. CURRENT SECURITY SOLUTIONS AND PRODUCTS – RELATED WORK

The concept of SOA changed the whole paradigm for the application development. As developers seem to adopt SOA anxiously, but at the same time, SOA implementation can lead to complex security and management issues. According to the survey conducted in 2008, 43% of the IT executives rated security threats as most critical issues while migrating towards SOA and 57% admits that security threats are one of the main reasons for adjourning the adoption of SOA. The majority of organizations migrated towards SOA are still at risk and putting a lot of efforts for securing their SOA environment. Now a day to secure SOA environment, developers are using current security approach and products provided by companies like IBM, HP, SOA Software,

Oracle and SAP. This section provides an overview of current security solution and products provided by various companies.

## 3.1 SOA Software Solution

SOA Software is one of the leading service providers in the field of SOA security and management. SOA Software developed a product named as an SOA Infrastructure Reference Model based on current standards [9]. The SOA Software's product contains the following:

### 3.1.1 Service Manager

It acts as policy manager and the main of service manager is to provide security services. Service manager also provides high performance and scalable SOA management solutions. Service manger includes management server and the management server helps to manage the various application monitoring functions such as registry and alert manager.

### 3.1.2 Network Director

Network Director designs and implements the policies for the service manager. The main objective of the Network Director is to provide fault tolerance capabilities, so that the available services can be consumed by the widest possible range of applications. It delivers comprehensive governance, security, monitoring, management and mediation of Web services.

### 3.1.3 Partner Manager

partner manager ensures that organizations can securely publish their web services for their partners and customers. It provides a complete trading partner management, and service provisioning workflow solution. It uses service virtualization to securely extend internal services into partners' networks.

### 3.1.4 SOLA

SOLA stands for Service Oriented Legacy Architecture (SOLA) and SOLA referred as complete mainframe Service Oriented Architecture solution [10]. The most important advantage of SOLA is that it makes the mainframe applications as a part of SOA in a cost effective manner. SOLA provides customers with a fast and easy process to expose mainframe applications as secure Web Services, and allows mainframe applications to consume Web Services. The SOLA runtime environment runs entirely on the mainframe, eliminating the need for expensive, unreliable and unnecessary middleware. SOLA is the only product proven in enterprise implementations to handle high volume (10 million+) transactions per day in mission-critical mainframe SOA environments. The various advantages of SOLA are comprehensive mainframe SOA solution, one click web services, no extra software's needed to install at work station and centralized directory.

## 3.2. IBM SOA Security Reference Model

Figure 3 shows the basic structure of IBM SOA Security Reference Model [11], this security model mainly divided into three different parts which are Business Security Services, IT Security Services and Security Policy Management. Business Security Services represent the main function of business security services is to manage the business requirement, for example identity management, data protection and securing networks. IT Security Services can be defined as the foundation wedges of an SOA infrastructure. IT Security Services makes the SOA environment capable of stopping up the services offered with the SOA infrastructure. In general, IT Security Services form a universal statement of services, which can be utilized by different users or components in the SOA environment. These services cover confidentiality, integrity, identity, authorization and authentication. Security Policy Management describes and design the various outlines related to the implementation of security policies in SOA environments. This includes the ability to define policies to authenticate and authorize requesters to access services, propagate security context across service requests based on an underlying trust model, audit events of significance, and protect information. Thus, Security Policy Management functionality is a core part of providing security capability in SOA. IBM SOA Security Reference Model was developed by IBM and the main objective behind this is to ascertain the security in SOA environments. To ascertain the security in a real world SOA environment, IBM also developed several products and instruments such as Tivoli and WebSphere. Tivoli and WebSphere are discussed next.

### 3.2.1 WebSphere Application Server

WebSphere Application Server is used for E-commerce and integration of organisational information. It can be defined as software framework used for hosting Java based applications. WebSphere Application Server promises to save the rich application development environment and it also provides superior transaction capabilities for transaction management. WebSphere Application Server has highly improved security model, as it is designed and produced on the basis of services provided in the operating organization. WebSphere Application Server supports user authorization and certification. The primary advantages of WebSphere Application Server are improved security and control standards, includes liberty Profile, superior capabilities for deployment and management for applications and higher developer productivity.

### 3.2.2 IBM Tivoli security management solutions

IBM Tivoli security management solutions designed to ensure the dynamic organizations and supply them a perfect entourage of security and compliance management. It solves the two critical problems for E-commerce, which is security management and identity management. It assists organizations in recovering from security attacks and system failures by bringing users and application online. It also assists organizations to actively monitor and react to security attacks. The primary benefits of IBM Tivoli solution are managing risk by specifying and deploying secure infrastructure, improved services and Reducing Cost of management, governance and mental process of the security infrastructure while increasing productivity.

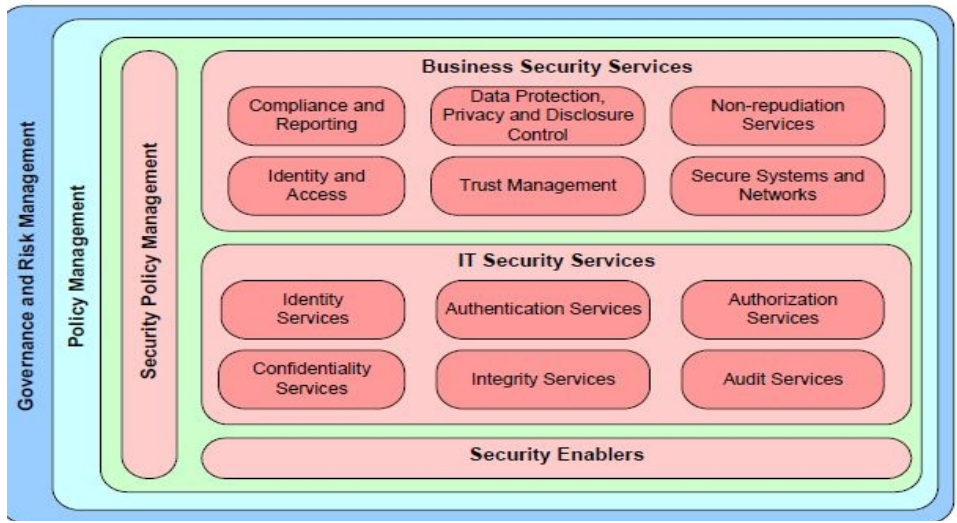

Figure 3 IBM SOA Security Reference Model

### 3.3 JBoss Enterprise SOA (ESOA) Platform

JBoss ESOA is open source software primarily based on Service Oriented Architecture and Java. It can be implemented independently or with other third party components as well. JBoss ESOA enables enterprises to properly handle business events, Integrate services; Automate existing business operations and linking existing IT resources with information and services. According to [12], "The versatility of the program permits us to use one development tool, technology and set of touchstones for many uses. Likewise, applications deployed on the JBoss ESOA Platform can be monitored and contained in a standardized way". JBoss itself is freely licensed but also called for other middleware components such as JBoss Enterprise Service Bus (ESB) and JBoss Enterprise Application Platform. JBoss can be applied with whatever operating system such as Windows, UNIX and Linux. The main implementation benefits of JBoss are improved business process performance, improved reliability and scalability, higher customer satisfaction and Eliminate Manual Pain Points from your Business Processes.

### 3.4 Vordel Solution

Vordel is an organization which provides various hardware and software solutions to implement SOA. The primary benefits of Vordel solution are Efficient security mechanism against various security attacks such as XML content-layer attacks, Improved monitoring of web serving performance and usage, Improved performance by removing processing bottlenecks, Enhanced governance through central service management, Well defined access control including authentication and authorization and Proper control for SOA policy.

### 3.5 Oracle SOA Security Solution

Oracle SOA Security Solution ensures the security outside the web serves as well. This solution can combine transport-layer and application-level protection, and use a layered defense scheme. The diverse benefits and products provided by Oracle SOA Security Solution are:

- Identity management includes Oracle Access Manager, Oracle Identity Federation and Oracle Web Services Manager
- Deployment and development tools include Oracle Security Developer Tools (OSDT)
- Java Developer consists of Java Integrated Development Environment
- Governance aware runtime environment comprises of Oracle components for Java (OC4J)

Oracle's strategy is to focus on well established criteria such as SAML, essential for identity federation, identity propagation, and close-to-end security from the user's web browser all the way across SOA-based transactions involving multiple Web Services.

## 4. SECURITY ANALYSIS OF SOA BASED E-COMMERCE

E-commerce provides user with a variety of business service components and user can call any of these business service components online such as trading information services, contract management services and other information management services, if they are authenticated and passed. Before calling business service component's user submits their authentication through HTTP to a WWW host. Users can log on to the component requested through Simple Object Access Protocol (SOAP) and at the same time user may utilize other services as well, which is furnished by other collaborators such as banking and logistics services. E-commerce platform is not hardly used for online transactions, but also can be practiced as a serial publication of services such as banking and revenue enhancement. So, E-commerce provides a complete lay down of business procedures which consist of various inspection and repairs. These services are originated by different vendors and managed or kept up by diverse departments within an endeavor, which makes E-commerce a service oriented heterogeneous system. This heterogeneous system will confront a kind of security threats and risk factors which are:

### 4.1 Certificate Duplicity

Approach to business service components, users should submit their authentication through HTTP and get it validated from web hosts. In return an authentication credential is released for users as a validation of certification. After getting this authentication certificate user can access business service components. This process of issuing authentication certificate is managed by Identity Management Service (IMS). It is possible to generate a duplicate certificate or forge the certificate and submit it to the web server instead of the original user and identity authentication is bypassed and call to the Business Service Component can be attained. One of the possible implementation solutions for this is to use better communication protocols such as HTTPS, which can't be worked well.

### 4.2 No filters mentioned on the application level

Filters are really important at application level and absence of these filters may allows an attacker to send malicious codes through web application and perform attacks like Cross Site Scripting, Remote/Local File Inclusion etc. which may immediately or indirectly lead to a successful attack.

### 4.3 Unsecure Database

In most of E-commerce models, database is maintained on the same server without going through any security bars. Thither are a number of security risks related to database that are as follows:

• Malware infections into the database may also go to incidents like leakage or disclosure of important information.

## 5. PROPOSED E-COMMERCE MODEL BASED ON SOA

The proposed Security Framework of SOA based E-commerce is shown in Figure 4. The proposed security framework resolves the security threats and risk involved in E-commerce security. In the proposed framework, the authentication of user is handled by Identity Management Service (IMS). IMS provides access to the authorized and authenticated users for various business service components such as Transaction Information Service and Online Transaction Services. Proposed framework also implemented on E-commerce systems which supports a series of services such as insurance services and banking services. User can call any of the supported services, by sending access request to that corresponding service through IMS. IMS in return generates a certificate of authentication for the users. The proposed security framework has the additional feature called as input sanitization. The main aim of the input sanitization is to filter the users request according to the services available in the database, if no corresponding services are available the users' requests denied. Rule based plug-in is defined in the proposed security framework to offer an extra layer of protection. Intrusion detection system (IDS) and Intrusion protection system (IPS) is employed in the proposed security framework to provide spare security to business service components and database related. In offering a security framework, business service components and connected database are maintained on different hosts. Server 1 maintains business service components and network hosts. Server 2 keeps up the databases connected with business service components. The communication between server 1 and server 2 is monitored by a firewall. If an attack detected by Intrusion detection system (IDS), firewall will disconnect the both servers from one another to assure minimal data loss.

## 6. IMPLEMENTING THE PROPOSED SOLUTION

The proposed security framework implemented on open source E-commerce system. Through which the proposed framework analyzed to discover its strengths and weakness in the light of security threats discussed above. The methodology for evaluating the framework depends upon the character of security threats. The rating also depends on the security requirements and arranging for the endeavour. Some of the character attributes which already achieved, such as database encryption shown in figure 5.

International Journal of Advanced Information Technology (IJAIT) Vol. 4, No. 3, June 2014

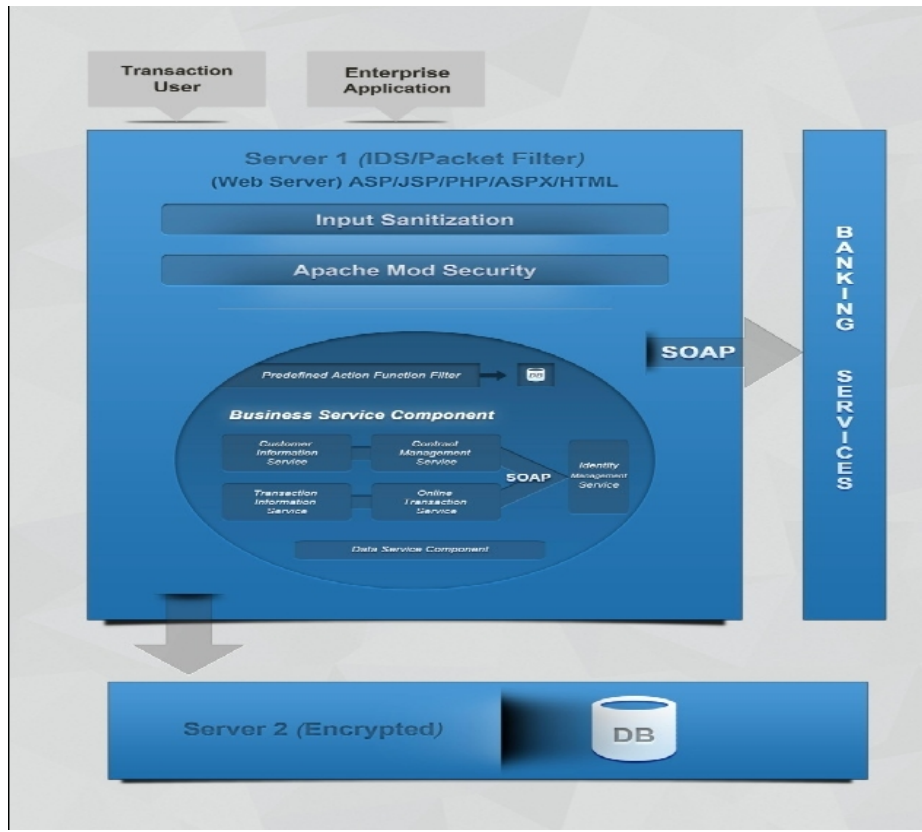

Figure 4 The Architectural overview of the proposed model

## 6.1 Database Encryption

The database maintained onto a totally different host (Server 2) which monitored by IDS/IPS. The database itself encrypted with a unique key which stored in data service component on server 1. Figure 5 shows the encrypted database for the proposed security design.

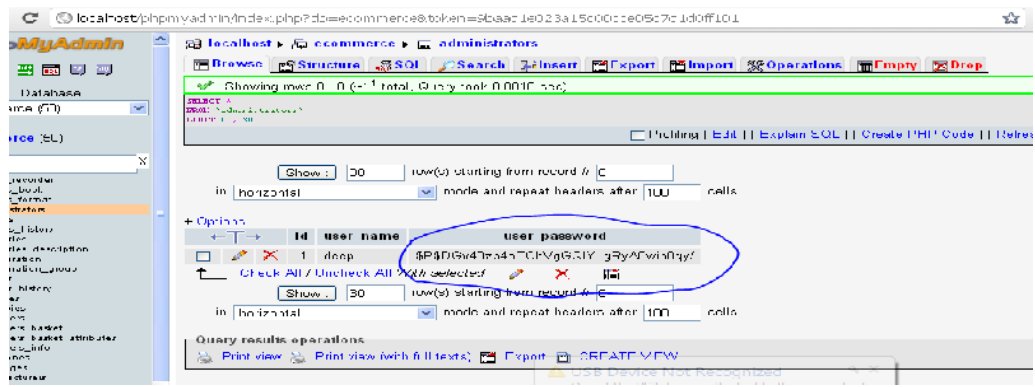

Figure 5 Encrypted Database

### 6.2 Prevention against DOS and DDOS

The presentation of IDS/IPS on both the servers will avoid most of the Denial of Service attacks as all the incoming data would be supervised and separated out by them. The IDS/IPS can be easily configured and automatically outlaw the IP Addresses generating multiple requests within a suitable interval of fourth dimension. Snort is open source software, which is used as IDS/IPS. Figure 6 shows how incoming data and activities can be monitored in proposed security design.

Figure 6: Snort as IDS

## 7. CONCLUSION

In today's rising information technology infrastructure, E-commerce is a promising business opportunity which continues and joins the business partners and customers in the electronic environment. The purpose of this research is to identify the best technology for E-commerce implementation, which is hardware and software platform neutral. This paper proposes an implementation scheme for E-commerce applications, which based on SOA. The proposed strategy has a logical security cover for user transactions and databases.

## REFERENCES:

[1] Min Huang et al, "Research for E-Commerce Platform Security Framework Based On SOA", proceeding of the 2011 4th International Conference on Biomedical Engineering and Informatics (BMEI), Shanghai, China, Pg. No. 2171-2174, October 15-17, 2011.
[2] Luhach Ashish Kr., Dwivedi Sanjay k. "Service-Oriented Architecture and Web Services Concepts, Technologies, and Tools" proceeding of the IEEE International Conference on Computational Intelligence and Computing Research (ICCIC-2011), Kanyakumari, India, Pg. No. 147-150, Dec 15-18, 2011.
[3] Erl, T., SOA Principles of Service Design. 1 Ed., Indiana: Prentice Hall, 2008.
[4] Giblin, T. I. el at., 'Web Services Security Configuration in a Service-Oriented Architecture'. WWW 2005, May 10-14, 2005 pp. 1120-1121. Chiba, Japan: ACM.
[5] Kuppuraju, S., Kumar, A., & Kumari, G. P. (2007). 'Case Study to Verify the Interoperability of a Service Oriented Architecture Stack'. 2007 IEEE International Conference on Services Computing (SCC 2007). IEEE Computer Society.
[6] Ramaro Kannergant, Prasad Chodavarapu. SOA Security in Action. Manning Publishing, 2006.
[7] ChristopherSteel Ramesh Nagappan, Ray La. Core Security Patterns. London: PrenticeHall, 2005.
[8] Ramarao Kanneganti, Prasad Chodavarapu, SOA Security, Manning Publications Co., USA, 2008. January.
[9] SOA Software, Inc. Seven steps to SOA, Los Angeles, white paper, 2006.

[10] SOA Software, Inc. SOA Infrastructure Reference Model, Los Angeles (U.S.A.), White paper, 2002.
[11] Axel Buecker et al., Understanding SOA Security Design and Implementation, 7 ed., Redbooks: IBM Corp., November 2007.
[12] SOA Magazine, Online, Available: http://soa.sys-con.com.
[13] Luhach Ashish Kr., Dwivedi Sanjay k., Jha C. K., Designing a logical Security Framework for E-commerce System based on Service Oriented Architectures, International journal of Soft Computing, Vol. –5, No.-2, Pg. No 1-10, May 2014.
[14] Mohammad H. Danesh et al, "A framework for process and performance management in service oriented virtual organisations", International Journal of Computer Information Systems and Industrial Management Applications, Vol. -5, Pg. No. 203-215, 2013.
[15] Deepu Raveendran et al, "A Study on Secure and Efficient Access Control Framework for SOA", International Journal of Computer Science and Telecommunications, Vol.-3, No.-6, Pg. No. 71-76, June 2012.
[16] Seyyed Mohammad Reza Farshchi et al., "Study of Security Issues on Traditional And New Generation of E-Commerce Model, proceeding of the 2011 International Conference on Software and Computer Applications, Kathmandu, Nepal, July 1-2, 2011, published by IPCSIT, Singapore, Vol.-9, Pg. No. 113- 117.